\documentclass[conference]{IEEEtran}
\IEEEoverridecommandlockouts
\usepackage{setspace}
\usepackage{adjustbox}
\graphicspath{ {./images/} }
\usepackage{epstopdf}
\usepackage{textcomp}
\usepackage{multirow}
\usepackage{xcolor}
\usepackage{tabularx}
\usepackage{dcolumn}
\usepackage{multicol}
\usepackage{subfigure}
\usepackage{amssymb}
\usepackage{float}
\usepackage{enumitem}
\usepackage{amsmath,amsfonts}
\usepackage{algorithmic}
\usepackage{algorithm}
\usepackage{array}
\usepackage[caption=false,font=normalsize,labelfont=sf,textfont=sf]{subfig}
\usepackage{textcomp}
\usepackage{stfloats}
\usepackage{url}
\usepackage{verbatim}
\usepackage{graphicx}
\usepackage{cite}
\begin{document}

\title{SNR-Adaptive Optimal Threshold Design for Energy Detection in Dynamic Spectrum Access}
\author{
\IEEEauthorblockN{
Sushila Dhaka\IEEEauthorrefmark{1},
Jane-Hwa Huang\IEEEauthorrefmark{2},
Chih-Min Yu\IEEEauthorrefmark{3},
Li-Chun~Wang\IEEEauthorrefmark{1}\IEEEauthorrefmark{4}
}

\IEEEauthorblockA{\IEEEauthorrefmark{1}
Department of Electrical and Computer Engineering,  
National Yang Ming Chiao Tung University, Hsinchu, Taiwan}

\IEEEauthorblockA{\IEEEauthorrefmark{2}
Department of Electrical Engineering,  
National Chi Nan University, Nantou County, Taiwan}

\IEEEauthorblockA{\IEEEauthorrefmark{3}
Department of Information and Computer Engineering,  
Chung Yuan Christian University, Taoyuan, Taiwan}

\IEEEauthorblockA{Emails: sushila.eed08g.ee08@nycu.edu.tw, jhhuang@ncnu.edu.tw, hankycm7@gmail.com}
\IEEEauthorblockA{\IEEEauthorrefmark{4}
Corresponding author: wang@nycu.edu.tw}
}

\maketitle
 \begin{abstract}
This paper proposes an SNR-adaptive optimal threshold design framework for energy detection in Dynamic Spectrum Access (DSA). Unlike conventional constant false-alarm rate (CFAR)-based schemes that determine the sensing threshold solely from a predefined false-alarm constraint, the proposed method directly minimizes the total probability of error by deriving a closed-form analytical solution. The threshold optimization problem is formulated as a quadratic expression whose coefficients explicitly characterize the effects of signal-to-noise ratio (SNR) and number of samples. This analytical structure enables adaptive threshold selection under heterogeneous SNR conditions without exhaustive numerical search. Simulation results demonstrate that the proposed approach reduces the error probability compared with fixed-threshold and detection-constrained schemes, particularly in low-SNR regimes. Furthermore, the impact of SNR and number of samples on detection performance is systematically analyzed, providing deeper insight into the trade-off between false alarm and missed detection.
The proposed framework improves sensing reliability and practical adaptability in dynamic spectrum access systems. It also establishes a foundation for secure cooperative spectrum sensing, including blockchain-assisted aggregation mechanisms.
\end{abstract}


\section{Introduction}
The rapid proliferation of wireless devices and bandwidth-intensive applications has led to a substantial increase in spectrum demand, even though many licensed frequency bands remain underutilized \cite{matinmikko2025spectrum}. Dynamic Spectrum Access (DSA) has therefore gained considerable attention as a means to improve spectrum efficiency by allowing secondary users (SUs) to opportunistically access licensed bands while ensuring adequate protection for primary users (PUs) \cite{jiang2021decentralized}.  
At the core of DSA lies reliable spectrum sensing. Sensing inaccuracies may either cause harmful interference to PUs or unnecessarily restrict spectrum access for SUs \cite{zhou2025spectrumfm}. Among various sensing techniques, energy detection (ED) is widely adopted due to its low implementation complexity and its independence from prior knowledge of the PU signal structure \cite{9598349}. Despite these advantages, the performance of ED is highly sensitive to the selection of the local decision threshold, which directly controls the trade-off between false alarms and missed detections \cite{10085685}.  

In conventional ED-based spectrum sensing \cite{urkowitz2005energy}, the decision threshold is typically determined using the constant false alarm rate (CFAR) criterion \cite{kun2009new}, where a predefined false alarm probability is maintained under the null hypothesis $H_0$. Although CFAR-based approaches are simple and analytically convenient \cite{wang2011energy}, they implicitly emphasize false alarm control rather than overall sensing reliability. In practical DSA environments, PU activity is often asymmetric and time-varying, resulting in unequal prior probabilities of $H_0$ and $H_1$. In addition, cooperative spectrum sensing involves spatially distributed SUs that experience heterogeneous SNR conditions. Under such scenarios, a fixed threshold designed solely to satisfy a false alarm constraint may not achieve optimal performance, particularly in low-SNR regimes where missed detections become more significant.  
With an increase in SNR, the separation between $H_0$ and $H_1$ increases. $H_1$ shifts right with an increase in SNR. The difference between $H_0$ and $H_1$ depends on the local threshold value along with the SNR. The separation between $H_0$ and $H_1$ confirms the detection.
However, detection performance is not governed by SNR alone; it also depends critically on the selected threshold. A threshold that is overly conservative may suppress false alarms but increase missed detections, whereas a lower threshold may improve detection at the expense of higher false alarms. Consequently, minimizing the total probability of error provides a more balanced and practically meaningful performance criterion than relying solely on CFAR-based design.  

While considerable research has addressed threshold design in ED-based sensing, direct minimization of the total probability of error under heterogeneous SNR conditions has received comparatively less attention. Moreover, in cooperative DSA systems, accurate local decisions alone are insufficient. Malicious or unreliable SUs may falsify sensing reports, leading to incorrect global decisions and degraded spectrum utilization \cite{9598349}. Ensuring secure and trustworthy aggregation of sensing outcomes is therefore equally important \cite{10085685}. Blockchain technology offers a decentralized and tamper-resistant infrastructure capable of enhancing transparency \cite{10260322} and accountability in distributed wireless environments \cite{10915542}. By securely recording sensing reports and incorporating reputation-aware mechanisms, blockchain can discourage malicious behavior and promote reliable participation among SUs.  

Several prior works have investigated threshold selection strategies for energy detection under CFAR constraints and Bayesian formulations \cite{kun2009new, wang2011energy}. Existing approaches primarily focus on controlling $P_f$ or optimizing detection performance under fixed statistical assumptions.
 
While numerical optimization and grid-based search methods can efficiently determine near-optimal thresholds with modern computational resources, these approaches generally do not yield explicit closed-form expressions for the total probability of error. Consequently, they provide limited analytical insight into the dependence of the optimal threshold on key system parameters, such as the signal-to-noise ratio (SNR) and the number of samples  $K$. This limitation motivates the development of a closed-form formulation that facilitates deeper theoretical understanding of threshold behavior and system performance. 

In contrast, this work derives a closed-form expression for $P_e$ and reformulates the threshold optimization problem into a quadratic structure whose coefficients and roots explicitly characterize the optimal operating point. This analytical formulation provides deeper insight into threshold adaptation across heterogeneous SNR regimes while avoiding exhaustive numerical search.

Such explicit analytical characterization facilitates low-complexity implementation \cite{Zeng2010} and provides theoretical insight that is not readily available in purely numerical optimization frameworks.
Motivated by these considerations, this paper proposed an analytical framework for minimizing the probability of error ($P_e$) through adaptive threshold selection in local spectrum sensing. A closed-form expression of the error function is derived, and the corresponding optimal threshold is obtained by solving the resulting quadratic formulation. The proposed framework is further extended toward secure cooperative operation through blockchain-assisted validation. Together, these contributions aim to improve sensing reliability while strengthening trust in dynamic spectrum access systems. 

The main contributions of this paper are summarized as follows:

\begin{itemize}
\item Closed-form error minimization: We directly minimize the Bayesian probability of error for energy detection and derive a closed-form quadratic expression for the optimal threshold, providing analytical insight into its dependence on system parameters.

\item Characterization of adaptive thresholds for SNR: The analytical formulation reveals the structural relationship among SNR, the number of samples, and the optimal threshold, allowing the selection of adaptive thresholds under heterogeneous conditions.

\item Performance and trade-off analysis: We systematically analyze the impact of SNR and the number of samples on detection performance and demonstrate a consistent reduction in error probability compared with conventional CFAR-based and fixed-threshold schemes.

\item Low-complexity implementation: The closed-form solution enables real-time threshold computation with minimal computational overhead for practical DSA systems.
\end{itemize}

The rest of this paper is organized as follows. Section 2 describes the system model. Section 3 presents the proposed optimal error-probability formulation. Section 4 provides the simulation results and corresponding analysis. Finally, Section 5 concludes the paper and outlines future research directions.

\section{System model}
In Cognitive Radio Network (CRN), Secondary Users (SUs) opportunistically access underutilized licensed spectrum bands owned by a Primary User (PU). 
Each SU performs energy detection \cite{10466663} to sense the presence or absence of the PU. If a channel is available (i.e., no PU activity), SUs can use it for transmission.
PUs have priority in using the spectrum, and their presence must be detected accurately by SUs to avoid interference.

\begin{figure}[H]
\centering
\includegraphics[height=7 cm,width=8.5 cm]{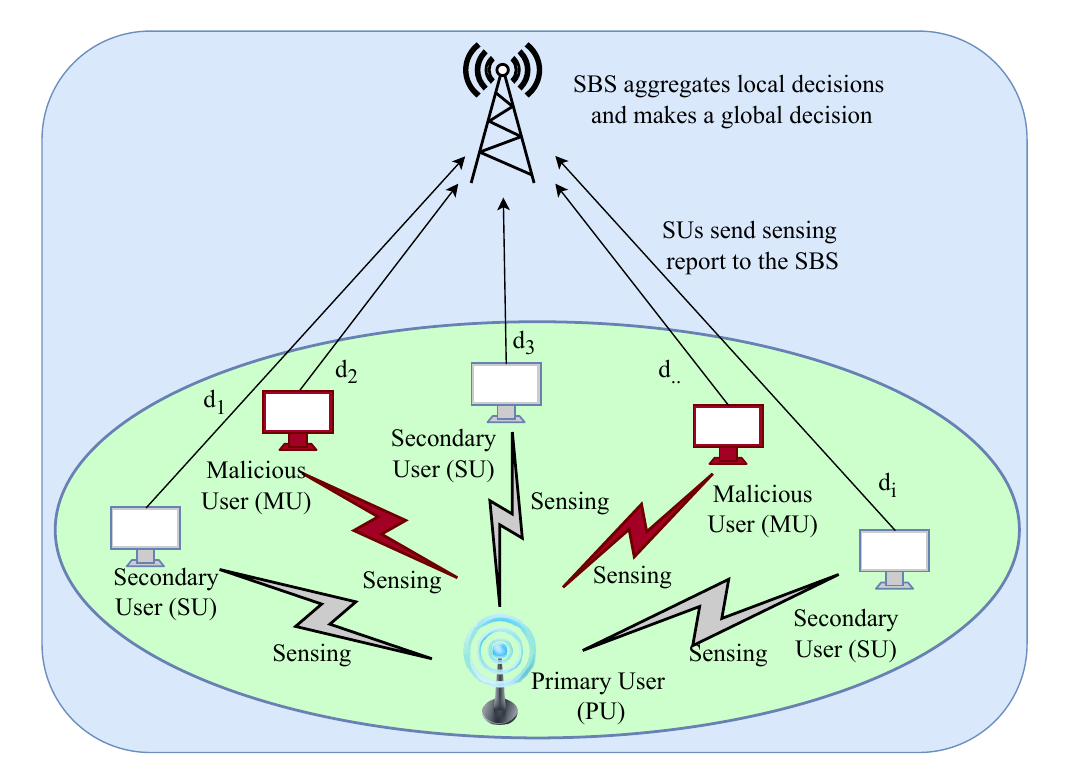}
\caption{System model for dynamic spectrum access.}
\label{fig:system model.}
\end{figure}

The integrity of spectrum sensing is susceptible to disruption by unreliable or malicious secondary users (MU). These SUs can submit false reports, leading to incorrect spectrum access decisions and potential interference with the PU.
This presents challenges in maintaining the correct spectrum status.
 To address these challenges, we propose an optimal probability of error.

\subsection{Energy Detection Method for PU Detection}

The energy detector measures the received signal energy at the $i$th secondary user (SU$_i$) and compares it with a decision threshold to determine the presence ($H_1$) or absence ($H_0$) of the primary user (PU). The received signal model is given by

\begin{equation}
\begin{cases}
H_0: \; y_i(t) = w(t), \\
H_1: \; y_i(t) = h_i s(t) + w(t),
\end{cases}
\label{eq:signal_model}
\end{equation}

where $w(t)$ denotes additive white Gaussian noise (AWGN), $s(t)$ is the transmitted PU signal, and $h_i$ is the channel gain of SU$_i$.

The energy collected over a sensing interval of $2K$ samples is defined as

\begin{equation}
\varepsilon_i = \sum_{t=1}^{2K} |y_i(t)|^2.
\label{eq:energy}
\end{equation}

For analytical tractability, the normalized energy statistic is defined as

\begin{equation}
E_i = \sum_{t=1}^{2K} \left| \frac{y_i(t)}{\sigma} \right|^2,
\label{eq:normalized_energy}
\end{equation}

where $\sigma^2$ is the noise variance.

Since $E_i$ is the sum of $2K$ independent squared Gaussian random variables, it follows a chi-square distribution. For sufficiently large $K$, the distribution of $E_i$ can be accurately approximated as Gaussian using the central limit theorem (CLT). Thus,

\begin{equation}
\begin{cases}
H_0: \; E_i \sim \mathcal{N}(\mu_{i0}, \sigma_{i0}^2), \\
H_1: \; E_i \sim \mathcal{N}(\mu_{i1}, \sigma_{i1}^2).
\end{cases}
\label{eq:gaussian_model}
\end{equation}

Under $H_0$, only noise is present, yielding
\[
\mu_{i0} = 2K,
\qquad
\sigma_{i0}^2 = 4K.
\]
Under $H_1$, the received signal contains both PU signal and noise. Assuming the PU signal is independent of the noise and defining the received SNR as $\gamma_i$, the mean and variance become

\[
\mu_{i1} = 2K(\gamma_i + 1),
\qquad
\sigma_{i1}^2 = 4K(2\gamma_i + 1).
\]

The local decision is obtained by comparing $E_i$ with a threshold $\lambda_i$:

\begin{equation}
d_i =
\begin{cases}
1, & E_i \ge \lambda_i, \\
0, & \text{otherwise}.
\end{cases}
\label{eq:decision_rule}
\end{equation}

Here, $d_i = 1$ indicates that the PU is present ($H_1$), while $d_i = 0$ indicates absence ($H_0$).

The probability of detection and probability of false alarm at SU$_i$ are given by

\begin{equation}
P_{d,i} = P(d_i = 1 \mid H_1)
= Q\!\left( \frac{\lambda_i - \mu_{i1}}{\sigma_{i1}} \right),
\label{eq:pd}
\end{equation}

\begin{equation}
P_{f,i} = P(d_i = 1 \mid H_0)
= Q\!\left( \frac{\lambda_i - \mu_{i0}}{\sigma_{i0}} \right),
\label{eq:pf}
\end{equation}

where $Q(\cdot)$ denotes the tail probability of the standard normal distribution.

The Bayesian probability of error is formulated as a linear combination of the false alarm probability and the miss-detection probability \cite{10598350}:

\begin{equation}
P_{e,i}
=
(1-\alpha) P_{f,i}
+
\alpha (1 - P_{d,i}).
\label{eq:pe}
\end{equation}

where $\alpha$ is the prior probability that the primary user (PU) is present.
The optimal threshold $\lambda_i$ is obtained by minimizing $P_{e,i}$.

\section{Proposed Optimal Probability of Error}
For a given SU$_i$, the objective is to determine the optimal threshold

\[
\lambda_{i,{opt}}=\arg\min_{\lambda_i} P_{e,i}(\lambda_i).
\]
From (\ref{eq:pe}), the probability of error at SU$_i$ is given by

\[
P_{e,i}(\lambda_i)
=
(1-\alpha)P_{f,i}(\lambda_i)
+
\alpha\bigl(1-P_{d,i}(\lambda_i)\bigr),
\]
where $P_{d,i}(\lambda_i)$ and $P_{f,i}(\lambda_i)$ are defined in (\ref{eq:pd}) and (\ref{eq:pf}), respectively.
Let

\[
x_{i0} \triangleq \frac{\lambda_i-\mu_{i0}}{\sigma_{i0}},
\qquad
x_{i1} \triangleq \frac{\lambda_i-\mu_{i1}}{\sigma_{i1}}.
\]
The standard normal probability density function (PDF) and the $Q$-function are

\[
\phi(x)=\frac{1}{\sqrt{2\pi}}e^{-x^2/2},
\qquad
Q(x)=\int_x^{\infty}\phi(t)\,dt.
\]
Taking derivatives, we use
\begin{equation}
\frac{dQ(x)}{dx}=-\phi(x).
\end{equation}
Applying the chain rule yields

\begin{equation}
\frac{d\,Q(x_{ik}(\lambda_i))}{d\lambda_i}
=
-\phi(x_{ik})\frac{dx_{ik}}{d\lambda_i},
\end{equation}
with

\[
\frac{dx_{i0}}{d\lambda_i}=\frac{1}{\sigma_{i0}},
\qquad
\frac{dx_{i1}}{d\lambda_i}=\frac{1}{\sigma_{i1}}.
\]
Differentiating $P_{e,i}(\lambda_i)$ with respect to $\lambda_i$ gives

\begin{equation}
\frac{dP_{e,i}}{d\lambda_i}
=
(1-\alpha)\!\left(-\frac{1}{\sigma_{i0}}\phi(x_{i0})\right)
+\alpha\!\left(\frac{1}{\sigma_{i1}}\phi(x_{i1})\right).
\label{eqn.11}
\end{equation}
Setting $\frac{dP_{e,i}}{d\lambda_i}=0$ yields

\begin{equation}
\frac{\alpha}{\sigma_{i1}}\phi(x_{i1})
=
\frac{1-\alpha}{\sigma_{i0}}\phi(x_{i0}).
\label{eqn.12}
\end{equation}
Substituting the value of $\phi(x)$ into \eqref{eqn.12} gives

\begin{equation}
\frac{\alpha}{\sigma_{i1}}\exp\!\left(-\frac{x_{i1}^2}{2}\right)
=
\frac{1-\alpha}{\sigma_{i0}}\exp\!\left(-\frac{x_{i0}^2}{2}\right).
\label{eqn.15}
\end{equation}

Taking the natural logarithm of both sides and substituting $x_{i0}$ and $x_{i1}$ yields

\begin{equation}
\frac{(\lambda_i-\mu_{i1})^2}{\sigma_{i1}^2}
-
\frac{(\lambda_i-\mu_{i0})^2}{\sigma_{i0}^2}
=
2\ln\!\left(\frac{\alpha\sigma_{i0}}{(1-\alpha)\sigma_{i1}}\right).
\label{eqn.16}
\end{equation}
Simplifying \eqref{eqn.16} leads to the quadratic equation

\begin{equation}
 \begin{split}
&(\sigma_{i0}^2-\sigma_{i1}^2)\lambda_i^2 + 2(\mu_{i0}\sigma_{i1}^2-\mu_{i1}\sigma_{i0}^2)\lambda_i \\ 
&\quad + (\mu_{i1}^2\sigma_{i0}^2-\mu_{i0}^2\sigma_{i1}^2) + 2\sigma_{i0}^2\sigma_{i1}^2 \ln\!\left(\frac{(1-\alpha)\sigma_{i1}}{\alpha\sigma_{i0}}\right) = 0.
 \end{split}
\label{eqn.17}
\end{equation}
Rewrite \eqref{eqn.17} in quadratic form as

\begin{equation}
A_i\lambda_i^2 + B_i\lambda_i + C_i = 0,
\label{eqn.18}
\end{equation}
where
\begin{align*}
A_i &= \sigma_{i0}^2-\sigma_{i1}^2, \\
B_i &= 2(\mu_{i0}\sigma_{i1}^2-\mu_{i1}\sigma_{i0}^2), \\
C_i &= \mu_{i1}^2\sigma_{i0}^2-\mu_{i0}^2\sigma_{i1}^2
+2\sigma_{i0}^2\sigma_{i1}^2
\ln\!\left(\frac{(1-\alpha)\sigma_{i1}}{\alpha\sigma_{i0}}\right).
\end{align*}
Using $\mu_{i0}=2K$, $\sigma_{i0}^2=4K$, $\mu_{i1}=2K(\gamma_i+1)$, and $\sigma_{i1}^2=4K(2\gamma_i+1)$, the coefficients reduce to

\begin{align*}
A_i &= -8K\gamma_i, \\
B_i &= 16K^2\gamma_i, \\
C_i &= 16K^3\gamma_i^2 + 32K^2(2\gamma_i+1) \ln\!\left(\frac{1-\alpha}{\alpha}\sqrt{2\gamma_i+1}\right).
\end{align*}
The discriminant

\begin{equation}
\Delta_i = B_i^2 - 4A_iC_i
\label{eqn.19}
\end{equation}
simplifies to
\[
\Delta_i
=
256K^3\gamma_i(2\gamma_i+1)
\Big[
K\gamma_i
+4\ln\!\left(\frac{1-\alpha}{\alpha}\right)
+2\ln(2\gamma_i+1)
\Big].
\]
Thus, the two candidate roots are

\begin{equation}
\lambda_{i,1,2}
=
\frac{-B_i \pm \sqrt{\Delta_i}}{2A_i}
=
K \mp \frac{\sqrt{\Delta_i}}{16K\gamma_i}.
\label{eqn.20}
\end{equation}
Since
\[\lambda_{i,1} = K-\frac{\sqrt{\Delta_i}}{16K\gamma_i}\]
is much less than $\mu_{i0}=2K$,
the false alarm probability at $\lambda_{i,1}$ will approach one, i.e. 
\[
P_{f,i}(\lambda_{i,1}) \approx 1,
\]
which is not a meaningful operating point. Therefore, the optimal threshold is chosen as

\begin{equation}
\lambda_{i, opt}
=
K+\frac{\sqrt{\Delta_i}}{16K\gamma_i}.
\label{eqn.21}
\end{equation}

\section{Results and Discussion}
All simulations were conducted using the Python software environment. The local decision threshold is a key parameter in the performance of the PU detection. A commonly adopted approach is the fixed $P_f$ (CFAR) technique, in which the threshold is chosen to satisfy a predefined false alarm probability and remains constant across SNR values. In contrast, the proposed analytical threshold $\lambda_{\text{opt}}$ is derived by minimizing the Bayesian error probability and is explicitly adapted to the SNR of each SU.

\begin{figure}
\centering
\includegraphics[height=6cm,width=7.5cm]{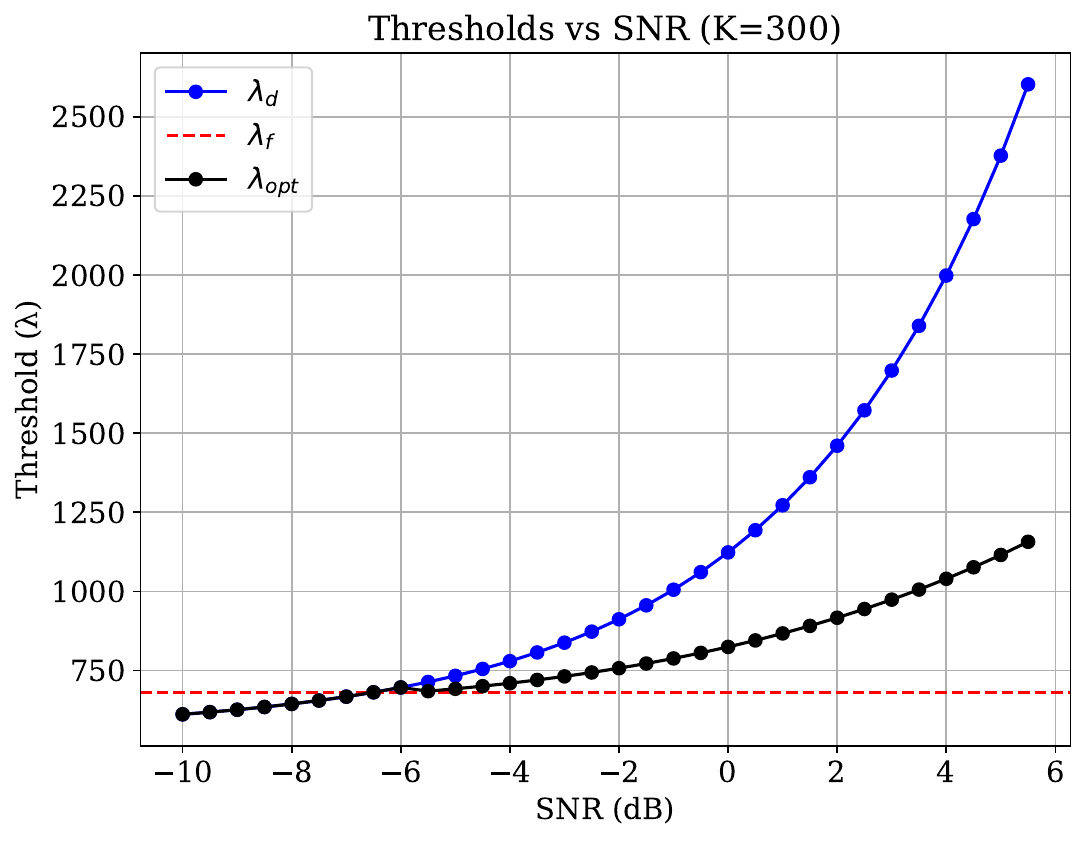}
\caption{Comparative analysis of SNR Vs decision threshold techniques.}
\label{fig: threshold.}
\end{figure}

Fig \ref{fig: threshold.} illustrates the threshold variation versus SNR. The fixed CFAR threshold $\lambda_f$ is constant and does not adapt to channel conditions. In contrast, $\lambda_d$ and the proposed $\lambda_{\text{opt}}$ vary with SNR. At low SNR, strong overlap between $H_0$ and $H_1$ requires a lower threshold, so both lie below $\lambda_f$. As SNR increases, the thresholds rise to suppress false alarms. The proposed $\lambda_{\text{opt}}$ minimizes the total error probability by optimally balancing missed detections and false alarms.

\begin{figure}
    \centering
    \includegraphics[height=6cm,width=7.5cm]{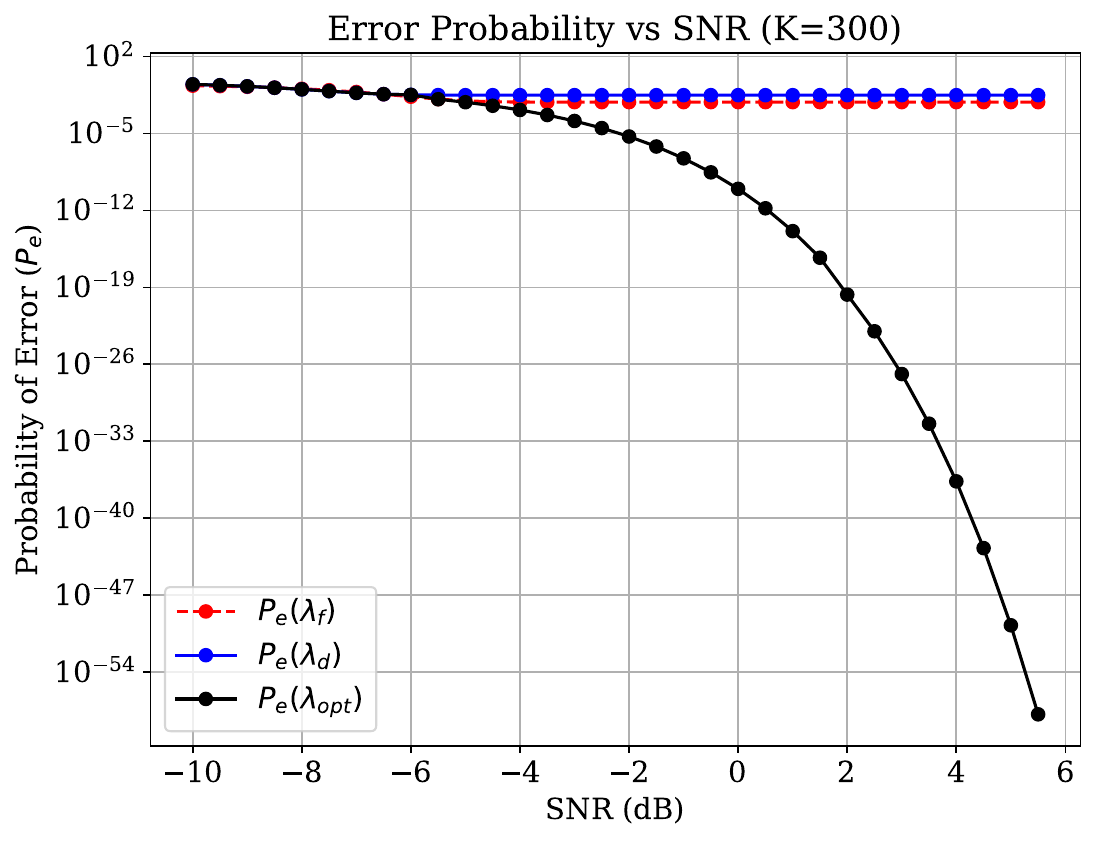}
    \caption{Error probability performance comparison under different threshold selection strategies.
}
    \label{fig:Pe_SNR}
\end{figure}

Fig \ref{fig:Pe_SNR} shows the error probability $P_e$ versus SNR for $\lambda_f$, $\lambda_d$, and the proposed $\lambda_{\text{opt}}$. The curve corresponding to $P_e(\lambda_f)$ appears nearly linear and remains essentially independent of SNR. For the detection-constrained threshold $\lambda_d$, $P_e$ varies only slightly because the imposed detection constraint effectively stabilizes the $Q$-function argument despite increases in mean and variance. In contrast, $P_e(\lambda_{\text{opt}})$ decreases rapidly with SNR, demonstrating that the proposed adaptive threshold achieves superior error minimization, particularly in medium and high-SNR regions.

\begin{figure}
    \centering
    \includegraphics[height=6.5cm,width=8cm]{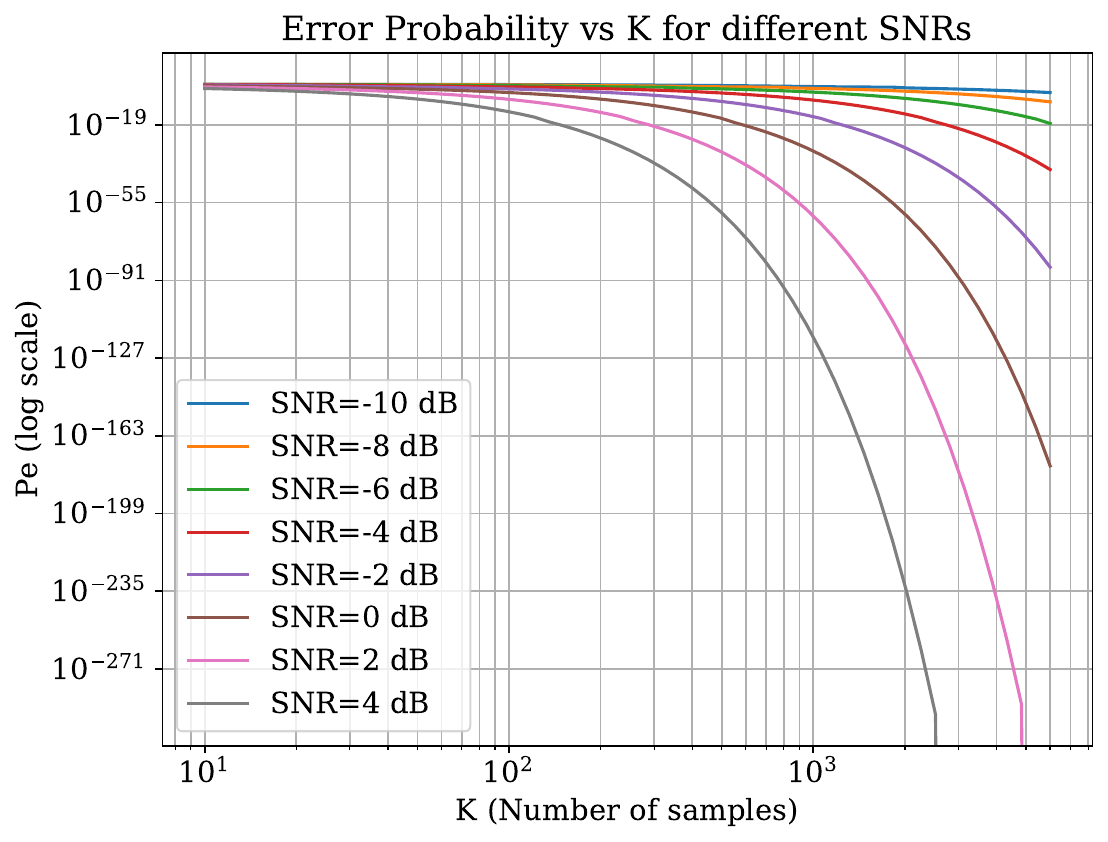}
    \caption{Error probability versus number of samples $K$ for different SNR levels.
}
    \label{fig:K_P}
\end{figure}

Fig \ref{fig:K_P} illustrates the effect of the number of samples $K$ on the error probability $P_e$ for different SNR levels. Since the test statistic is formed by summing $2K$ samples, performance depends jointly on both SNR and the integration length $K$. The variation of $P_e$ is therefore strongly influenced by these two parameters.
For small values of $K$, $P_e$ remains relatively high due to significant noise influence. As $K$ increases, averaging over a larger number of samples reduces noise fluctuations and enhances detection reliability, resulting in a decrease in $P_e$. 

At low SNR, this reduction is gradual because stronger noise dominance requires longer integration to achieve meaningful separation between $H_0$ and $H_1$. In contrast, at high SNR, $P_e$ decreases rapidly even for moderate values of $K$, since the stronger signal enables reliable detection with shorter integration lengths. This produces a steeper slope and faster decay in the high-SNR curves compared to the low-SNR cases.
Overall, the relationship among $P_e$, SNR, and $K$ reflects the fundamental trade-off in energy detection: increasing $K$ mitigates noise effects through averaging, while higher SNR inherently improves hypothesis separability, leading to the observed trends.

\begin{figure}
    \centering
    \includegraphics[height=6.5cm,width=8cm]{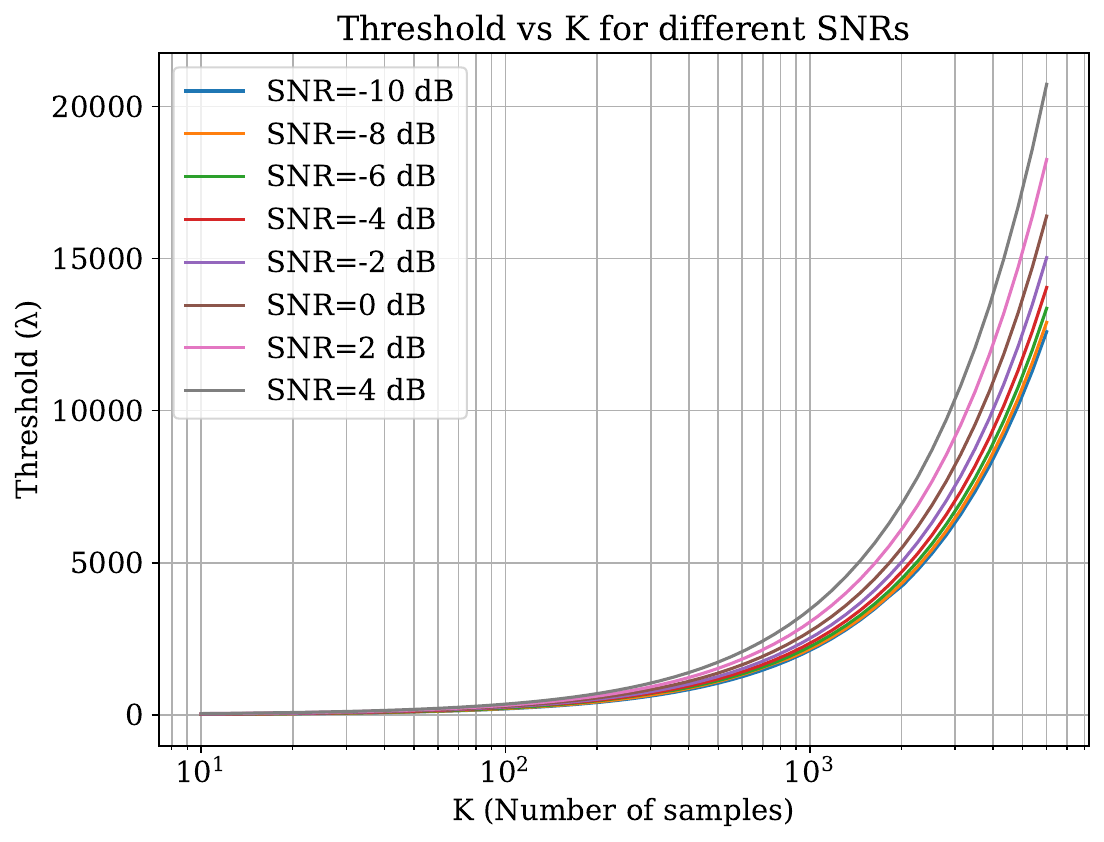}
    \caption{Optimal threshold $\lambda_{\text{opt}}$ versus number of samples $K$ for different SNR levels.
}
    \label{fig:K_lambda}
\end{figure}

Fig~\ref{fig:K_lambda} shows the variation of the optimal threshold $\lambda_{\text{opt}}$ with respect to the number of samples $K$ for different SNR levels. 
At low SNR, the system operates in a noise-limited regime where the distributions under $H_0$ and $H_1$ significantly overlap. In this case, the false alarm constraint dominates, and $\lambda_{\text{opt}}$ remains close to the fixed CFAR threshold $\lambda_f$.
As SNR increases, the separation between the means under $H_0$ and $H_1$ grows proportionally with $\gamma$, while the variance scales with $K$. Consequently, $\lambda_{\text{opt}}$ gradually shifts toward the detection-constrained threshold $\lambda_d$, reflecting the trade-off between minimizing $P_f$ and maximizing $P_d$.
At high SNR, the hypotheses become well separated, and the optimal decision boundary approaches $\lambda_d$, where the error probability is minimized. Increasing $K$ further enhances statistical averaging, amplifying this separation and resulting in the observed growth of $\lambda_{\text{opt}}$ with $K$. This behavior confirms the transition from noise-limited to signal-limited operation as SNR increases.

\begin{figure}
    \centering
    \includegraphics[height=6.5cm,width=8cm]{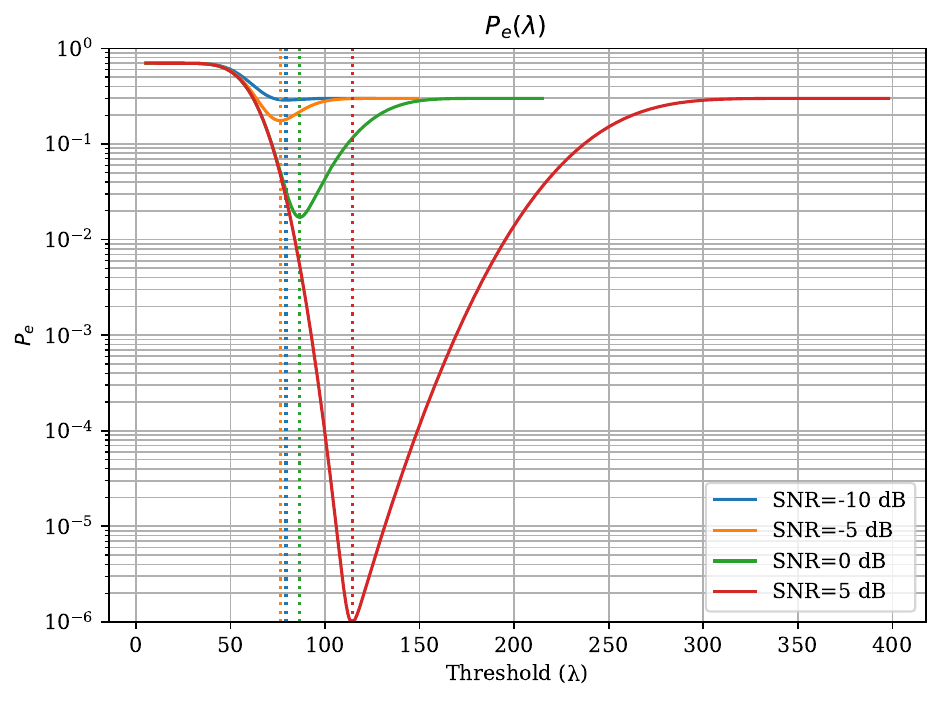}
    \caption{Error probability $P_e$ versus decision threshold for different SNR levels.
}
    \label{fig:Pe_opt}
\end{figure}

Fig \ref{fig:Pe_opt} presents the error probability $P_e(\lambda)$ as a function of the decision threshold for different SNR levels. The vertical dotted lines indicate the analytically derived optimal thresholds $\lambda_{\text{opt}}$.
At low SNR, the error surface exhibits a broad minimum, and small visual deviations between the sampled minimum and the analytical threshold may appear due to numerical resolution. As SNR increases, the curvature around the minimum becomes steeper, and the analytical $\lambda_{\text{opt}}$ aligns precisely with the minimum of the error curve. The increasing sharpness of the minima confirms improved hypothesis separability and validates the correctness of the closed-form threshold expression.

\section{Conclusion}

This paper presented a closed-form SNR-adaptive optimal threshold design framework for energy detection in dynamic spectrum access systems. By directly minimizing the Bayesian probability of error, the threshold optimization problem was reformulated into a quadratic expression, yielding an explicit analytical solution. Unlike conventional CFAR-based approaches that rely on predefined false alarm constraints, the proposed framework revealed the structural relationship among SNR, number of samples, and the optimal decision threshold.
The analytical formulation provided deeper insight into the trade-off between false alarm and missed detection probabilities across heterogeneous sensing conditions. Simulation results demonstrated consistent reduction in total error probability compared with fixed-threshold and CFAR-based schemes, particularly in low-SNR regimes. Owing to its closed-form structure and low computational complexity, the proposed method was well suited for practical and real-time dynamic spectrum access implementations. The proposed SNR-adaptive thresholding framework is particularly relevant for AI-driven and vehicular communication scenarios, where rapidly varying channel conditions and stringent latency constraints require analytically tractable and adaptive sensing mechanisms.

Future work will extend the proposed framework to cooperative spectrum sensing scenarios, where distributed users jointly optimize sensing thresholds under heterogeneous SNR conditions. In addition, integration with adaptive resource allocation and secure spectrum sharing mechanisms will be investigated to enhance robustness in large-scale and dynamic wireless environments. In particular, blockchain-enabled aggregation and validation mechanisms will be explored to ensure integrity, transparency, and trust in dynamic spectrum access networks.

\section*{Acknowledgment}
This work has been partially funded by the National Science and Technology Council under the NSTC 114-2221-E-A49-185-MY3, Grant, and NSTC 114-2221-E-033-026.
This work was supported by the Higher Education Sprout Project of the National Yang Ming Chiao Tung University and the Ministry of Education (MOE), Taiwan.

%*********************************************************************
%                               Reference
%*********************************************************************
\bibliographystyle{IEEEtran}

%*********************************************************************
\bibliography{reference}                              
%*********************************************************************

%\newpage
\vskip 0pt plus -1fil
%\vspace{-1cm}
%\begin{IEEEbiography}[{\includegraphics[width=1in,height=1.1in,clip,keepaspectratio]{Sushila.pdf}}]{Sushila Dhaka}

%\end{IEEEbiography}

\vskip 0pt plus -1fil

%\end{IEEEbiography}

\vskip 0pt plus -1fil
%\vspace{-1cm}

\vskip 0pt plus -1fil
\vspace{-1cm}

\end{document}